\documentclass{article}

\usepackage{arxiv}

\usepackage[utf8]{inputenc}            
\usepackage[T1]{fontenc}               
\usepackage[hidelinks]{hyperref}       
\usepackage{url}                       
\usepackage{booktabs}                  
\usepackage{amsfonts}                  
\usepackage{nicefrac}                  
\usepackage{microtype}                 
\usepackage{lipsum}		               
\usepackage{amsmath}
\usepackage{graphicx}
\usepackage{tabularx}
\usepackage{multirow}
\usepackage{array}
\newcolumntype{C}{>{\centering\arraybackslash}X}   
\newcolumntype{L}{>{\raggedright\arraybackslash}X} 

\usepackage[numbers,sort&compress]{natbib}
\usepackage{doi}
\usepackage{caption}

\title{From Sustainable Materials to User-Centered Sustainability: Material Experience in Art Healing}

    \author{ 
    \includegraphics[scale=0.06]{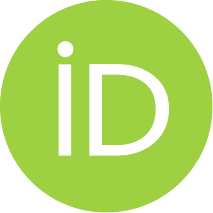}\hspace{1mm}Yuxin Zhang
        \thanks{Yuxin Zhang, PhD in Design, focuses on industrial design, material experience, affective engineering, and ergonomics. For further academic details, please refer to his ORCID profile: https://orcid.org/0000-0001-8100-3649} \\
        Academy of Arts \& Design \\
        Tsinghua University \\
        Haidian  District, Beijing, 100084, China \\
        \texttt{zhangyuxin@mail.tsinghua.edu.cn} \\
    \And
	Fan Zhang \\
	   Academy of Arts \& Design \\
	   Tsinghua University \\
	   Haidian  District, Beijing, 100084, China \\
	   \texttt{zhang-f24@mails.tsinghua.edu.cn} \\
    \And
	Zihao Song \\
        College of Materials Science and Engineering \\
        Beijing University of Chemical Technology \\
        Chaoyang District, Beijing, 100029, China \\
        \texttt{2021400176@buct.edu.cn} \\
    \And
	Chao Zhao \\
        Academy of Arts \& Design\\
        Tsinghua University\\
        Haidian District, Beijing, 100084, China \\
        \texttt{zhaochao@tsinghua.edu.cn} \\
}

\renewcommand{\shorttitle}{MATERIAL EXPERIENCE IN ART HEALING}

\hypersetup{
pdftitle={A template for the arxiv style},
pdfsubject={q-bio.NC, q-bio.QM},
pdfauthor={David S.~Hippocampus, Elias D.~Striatum},
pdfkeywords={First keyword, Second keyword, More},
}

\begin{document}
\maketitle

\begin{abstract}
This study develops sustainable materials using hydrogel as the matrix and explores the transition from sustainable materials to user-centered sustainability, with a particular focus on achieving art healing through material experience. The findings reveal that "Aesthetic" property exert the greatest influence on art healing in the context of multimodal material experiences involving visual, tactile, and smell, followed by "Intrinsic" property, whereas "Physical" property have a comparatively limited effect. Furthermore, the study proposes a material experience framework that enables designers to systematically and holistically understanding material characteristics. It highlights the importance of considering users' psychological perceptions and emotional needs in the material design process.
\end{abstract}

\keywords{Sustainable Materials \and Material Experience \and Computational Aesthetics \and Kansei Engineering}

\section{Introduction}
In a world confronted with limited resources and severe environmental challenges, the pursuit of a more sustainable lifestyle has become increasingly imperative \cite{ljungbergMaterialsSelectionDesign2007}. However, sustainability is a broad and multifaceted concept. From an ecological perspective, sustainability describes the ability of biological systems to remain healthy, diverse, and productive over time. Thus, sustainability is a concept that applies to higher levels of biological organization, such as ecosystems, rather than to individual species \cite{stuart-foxChallengesOpportunitiesInnovation2023}. From a human perspective, sustainability, or more specifically sustainable development, has been defined as "meeting the needs of the present without compromising the ability of future generations to meet their own needs" \cite{worldcommissiononenvironmentanddevelopmentReportWorldCommission1987}. Therefore, the achievement of the sustainable development goals (SDGs) is deeply intertwined with many areas of research and must remain at the forefront of future scientific and technological advancements. Given the complexity and multidimensional nature of the SDGs, achieving these goals often requires interdisciplinary collaboration and integrated approaches across multiple domains \cite{bontempiSustainableMaterialsTheir2021}. Among the disciplines contributing to sustainable development, design plays a pivotal role by shaping not only the functionality but also the emotional resonance of products \cite{baldassarreResponsibleDesignThinking2024, rossiResearchSynergiesSustainability2023}. Within this context, material selection emerges as a critical factor in bridging technical performance and user experience. As a key stage in the product design process, material selection significantly impacts the sustainability of manufacturing outcomes and simultaneously serves as a medium for communicating information and evoking emotional responses \cite{zarandiMaterialSelectionMethodology2011}. In many cases, this medium of communication proves to be even more important than the content it conveys \cite{mcluhanUnderstandingMediaExtensions1964}. However, selecting the appropriate material is a challenging process that requires managing a vast amount of information about material properties \cite{chinerPlanningExpertSystems1988}. Furthermore, only a limited number of design engineers possess a comprehensive understanding of all the characteristics of materials used in product manufacturing \cite{zarandiMaterialSelectionMethodology2011}. Therefore, a systematic and framework-based understanding of the role materials play in both ecological and social dimensions becomes particularly essential.

A considerable body of research has examined material sustainability, primarily focusing on the environmental impacts of materials and the recycling of resources \cite{hopewellPlasticsRecyclingChallenges2009, kumarUtilizationPlasticWastes2021, zhuSustainablePolymersRenewable2016}. In parallel, since the 20th century, the integration of art into healthcare has progressively expanded, encompassing fields such as architecture, interior design, creative arts therapies, and healing arts \cite{heineyHealingCreatingPatient2017}. An increasing number of studies has explored the effects of creative arts \cite{svenskArtTherapyImproves2009, woodWhatResearchEvidence2011}, which have been shown to alleviate anxiety, depression, and pain, while enhancing overall quality of life \cite{puetzEffectsCreativeArts2013}. Although the concept of "healing" is difficult to define \cite{rahtzUnderstandingPublicPerceptions2019}, within the context of complementary and alternative medicine (CAM), it is used to describe a wider phenomenon—including interventions, processes, outcomes, or psychological states \cite{levinWhatHealingReflections2017}. As a design discipline that emphasizes creative expression and sensory experience, material design aligns closely with the principles and practices of art healing. Another study also indicates that materials do not exist in isolation, but are closely intertwined with the natural environment, artistic expression, and healing logic. Through the interaction between people and materials, healing effects can be triggered. On a sensory level, materials activate bodily perception, rebuilding the mind-body connection; on a symbolic level, materials carry emotions, facilitating the transformation of trauma and emotions \cite{atkinsNaturebasedExpressiveArts2018}. A review conducted by Sonke et al. \cite{sonkeStateArtsHealthcare2009} on the application of arts in healthcare in the United States highlighted the absence of robust theoretical frameworks in earlier studies and underscored the need to clearly differentiate between empirical research and process-oriented evaluations. Moreover, relatively few studies have investigated patients' subjective perceptions regarding the benefits of participating in or receiving arts healing interventions \cite{heineyHealingCreatingPatient2017}. Against this backdrop, developing and designing materials from an artistic perspective—employing creative thinking to reconcile engineering performance with artistic expression—emerges as a significant and timely area of research. In recent years, the growing interest in art healing has encouraged scholars to reexamine its role in design, particularly from a user-centered perspective. Notably, Karana et al. \cite{karanaMaterialsExperience2015} have introduced the concept of "Material Experience" as a way to bridge materialistic and perception, providing valuable insights into how materials can be examined through perceptual and experiential dimensions. Building on this perspective, we argue that material experience represents critical pathway for linking sustainable materials with user-centered sustainability. Specifically, in the context of art healing, the core of material experience lies in aligning sustainable behaviors with users' habits and experience expectations, rather than requiring users to compromise for "Environmental Protection". This allows users to effectively realize the sustainable value of materials while enjoying sensory experiences. At the same time, material experience also increases user engagement, enhancing long-term user loyalty and interaction frequency.

\begin{figure}[h]
\centering
\includegraphics[width=0.9\textwidth]{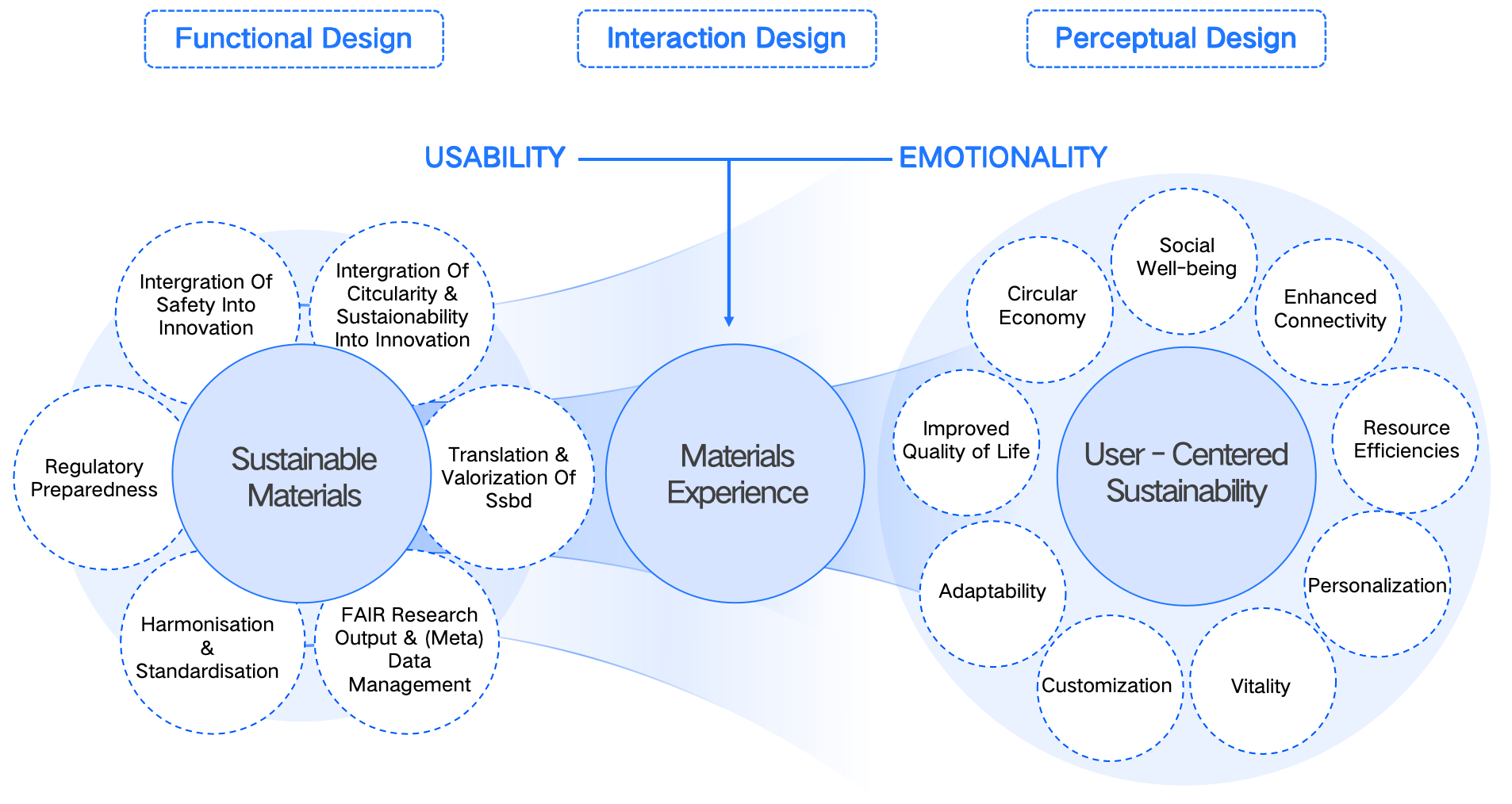}
\caption{A conceptual Framework Based on Material Experience: The Evolution from Sustainable Materials to User-Centered Sustainability}
\label{fig:1}
\end{figure}

However, the transition from sustainable materials to user-centered sustainability still faces numerous challenges and limitations. A typical example is eco-friendly straws: although they achieve environmental benefits and support resource recycling in terms of material selection and production, they often fall short in functionality and user experience during actual use. These shortcomings reduce user acceptance and usage frequency, ultimately limiting the widespread adoption of eco-friendly straws. This highlights that prioritizing material sustainability alone, while neglecting user needs and experiences, makes it difficult to realize truly meaningful sustainability. Therefore, sustainability should not only concern the materials themselves but must also embrace user-centered sustainable development. By enhancing the user experience during interactions with materials, and thereby increasing product usage frequency, the value of materials can be maximized, truly achieving the transition from sustainable materials to user-centered sustainability \cite{selveforsUsercenteredCircularValue2024, strappiniSustainableMaterialsLinking2024}. Figure \ref{fig:1} summarizes the transformation pathway from sustainable materials to human-centered sustainability \cite{casseeRoadmapSafeSustainable2024, hasaniOutlookHumancentredDesign2025, vanremmenSystematicLiteratureAnalysis2025}. We found that in the process of material design and selection, it is essential to focus not only on functionality but also on the perceptual dimensions of the materials. Only through a holistic and systematic approach to material planning can products be created that are not only technically functional but also emotionally resonant with users \cite{mcdonaghVisualProductEvaluation2002, suzukiFrameworkMaterialsPlanning2005}.

This study aims to explore the art healing potential of sustainable materials through the material experience, and to further leverage this potential to extend their application toward user-centered sustainable design. By integrating multisensory experiences, the study seeks to provide a new perspective for material design and promote a shift in sustainability thinking—from a traditional environmental focus to a more user-centered approach to sustainable development.

\section{Experiment}
\subsection{Method}
This study used a mixed-methods approach, combining quantitative analysis and qualitative questionnaires, to investigate the potential of sustainable materials in promoting art healing. Hydrogel was used as the matrix material and various sustainable fillers were added to create materials with different sensory properties. The sensory properties of these materials were treated as research variables during the material experience. Participants were engaged with different materials and provided subjective responses related to art healing as a critical indicator to evaluate the extent to which user-centered sustainability has been achieved. To systematically explore the impact of material experience on user perception, this study adopts the material experience framework proposed by Karana et al \cite{karanaMaterialDrivenDesign2015}. Using this framework, the relationship between material experience and art healing is further explored in relation to the purpose of this study. Then, a structural equation model is constructed to investigate the pathway from sustainable materials to user-centered sustainable development within the context of material experience. Additionally, the study integrates path coefficients and questionnaire data to further analyses how specific material characteristics, through the material experience mechanism, can guide sustainable materials towards achieving user-centered sustainable development.

\subsection{Stimulus}
\begin{figure}
\centering
\includegraphics[width=0.9\textwidth]{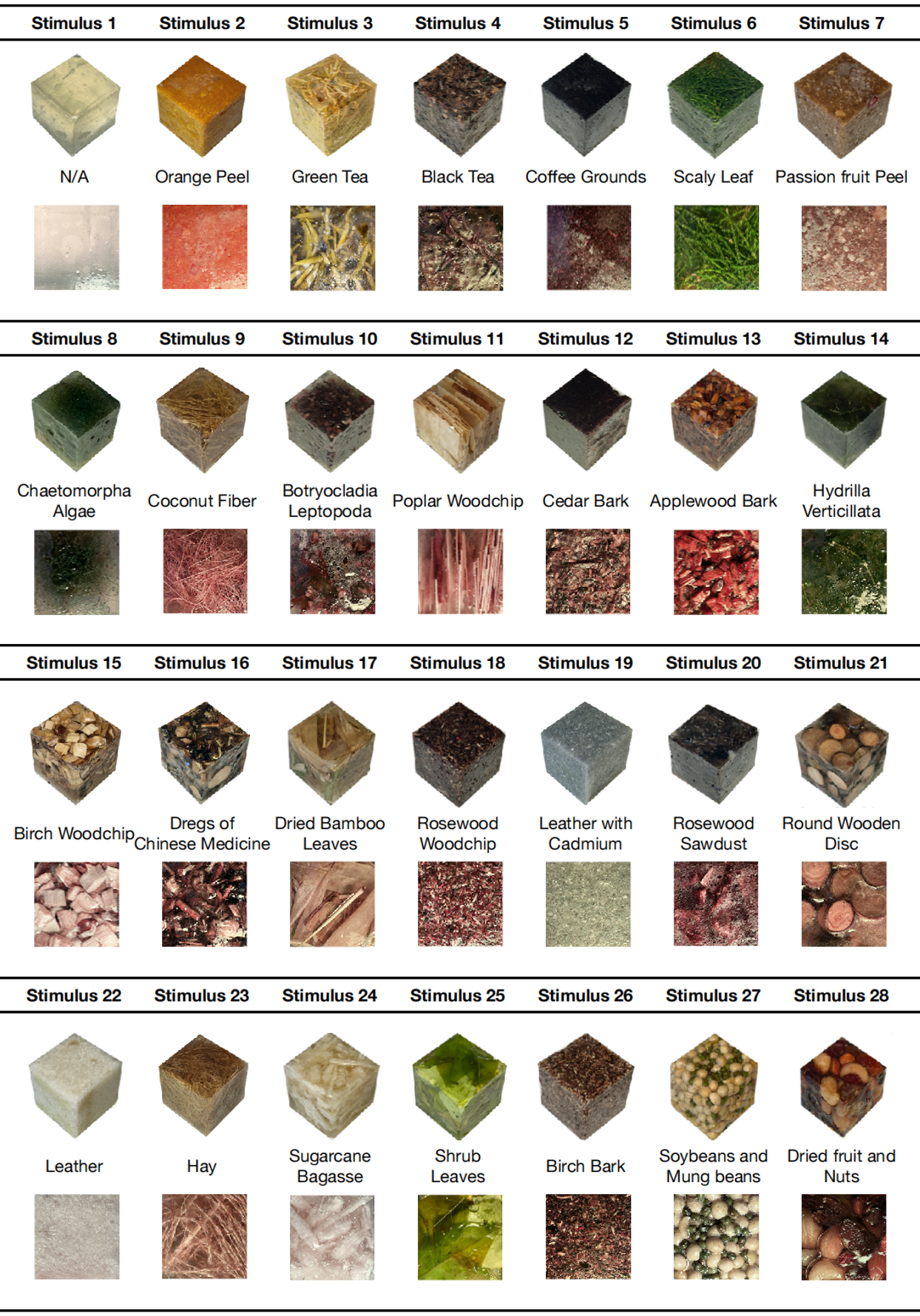}
\caption{Sustainable Material Design for Art Healing. Sustainable materials are prepared by adding sustainable fillers to hydrogel matrix. A total of 28 sustainable materials, including one non-filler only matrix and 27 composites were prepared}
\label{fig:2}
\end{figure}

The method of manufacturing the Stimulus is to compound the hydrogel matrix with sustainable filler. These sustainable fillers are mainly derived from natural material residues, natural organisms, and waste products with recycling value. Within the context of art healing, we emphasize the natural properties embedded in these fillers, exploring whether they can provide a unique sense of healing during material experiences \cite{elkis-abuhoffExploringEffectsNaturebased2022}. Among these materials, plant-based fillers have been included in the research due to their ability to achieve healing goals through "sensory interaction" and "ecological connection" \cite{gulbeExploringNaturebasedArt2025}. Additionally, we focus on animal leather as a bio-based material, known for its unique texture and durable physical properties. All of these fillers are derived from the natural environment, and it is this vitality sourced from nature that distinguishes them from traditional industrial materials. Through adding these sustainable fillers to the hydrogel matrix, a series stimulus was produced. A total of 28 stimuli were prepared, one of which was a hydrogel matrix without any filler as a control and the remaining 27 were composites of hydrogel matrix and sustainable filler. The prepared stimuli and the sustainable filler are shown in Figure \ref{fig:2}.

\subsubsection{Matrix Material}
The hydrogel matrix was prepared by combining gelatin and k-carrageenan. Specifically, gelatin (10 wt\%) was allowed to swell in deionized water (D.I. water) at room temperature for 30 minutes, followed by complete dissolution using magnetic stirring at 60 °C for 30 minutes. Next, k-carrageenan (2 wt\%) was added to the gelatin solution and continuously stirred for 1 hour to ensure homogeneity. The resulting solution was then poured into a 5 cm³ silicone mold and refrigerated at 2 °C for 2 hours to complete the gelation process. After demolding, transparent gel cubes with precise dimensions were obtained. For hydrogel composites containing filler materials, bulk fillers were carefully positioned in the silicone mold before casting the solution. Particulate or powdered fillers were first dispersed evenly in the hydrogel solution through homogenization, and then cast into the mold.

\subsubsection{Filler Material}
The filler materials selected for this study have sustainable properties. These biofillers present diverse fundamental characteristics that are expected to replace carbon black or inorganic mineral fillers, thereby reducing dependence on petroleum. In addition, they are biodegradable, which helps to effectively reduce the environmental impact \cite{dadkhahComprehensiveOverviewConventional2024}. Specifically, the filler materials for stimuli 2, 7, and 24 are derived from orange peel, passion fruit peel, and sugarcane bagasse, which are inedible parts of agricultural crops. Stimuli 8 and 14 use natural seaweed as a filler, whose unique structure provides excellent moisture absorption and moisturising. Additionally, the elasticity and softness of seaweed enable it to offer good flexibility and durability to certain materials, especially in applications that require enhanced structural stability. Stimuli 3, 4, and 5 use discarded green tea, black tea, and coffee grounds as fillers, all of which are waste products from commonly consumed beverages. We retain their original state and directly use them as fillers. The filler materials for stimuli 19 and 22 are leather with cadmium and leather. As a natural material, they offer exceptional durability and softness. Stimuli 9 and 23 use coconut fiber and hay, which have the characteristics of fibrous materials. We focus on how their fiber characteristics influence the perceptual experience of the materials. Stimuli 6 ,17 and 25 use scaly leaves, dried bamboo leaves and shrub leaves as fillers. As a natural material, plant leaves have unique textures, structures, and colours. Stimulus 16 uses traditional Chinese medicine residue, which is the by-product of Eastern herbal medicine after use. This residue typically comes from herbs that have been boiled multiple times and emits a unique odour, often with a strong herbal fragrance. In addition, Chinese medicine residue carries significant cultural value. Stimuli 27 and 28 use fillers composed of two mixtures: soybeans and mung beans, and dried fruit and nuts, respectively. We focus on the performance of non-single-form fillers in impression evaluation. The remaining stimuli (Stimuli 10, 11, 12, 13, 15, 18, 20, 21, 26) use wood as the filler, but we have controlled the form of the wood. This includes grinding it into fine particles (Stimuli 12, 18, 26), coarse particles (Stimuli 10, 13), slicing (Stimuli 11), cutting into segments (Stimuli 15, 20), and using it in its natural structure (Stimuli 21) as a filler. Based on the 27 sustainable fillers mentioned above, 27 sustainable composite materials were prepared.

\subsection{Impression Evaluation Item}
Having a detailed and reliable vocabulary is essential for accurately describing perceptual experiences \cite{guestDevelopmentValidationSensory2011a}. Therefore, we referred to studies on material experience involving visual, tactile, and smell perception \cite{sakamotoExploringTactilePerceptual2017, zhangIMPRESSIONEVALUATIONBASED2024, zuoSensoryPerceptionMaterial2016}, and selected 15 frequently mentioned pairs of impression evaluation items from the literature to ensure the generalizability of the study. In addition, a pre-experiment was conducted with five participants.

Based on visual, tactile, and smell interaction, participants freely experienced the stimuli and described the material experiences using positive and negative meaning adjectives. This process helped ensure that the selected impression evaluation items were sufficiently relevant to the stimuli in this study. We collected 135 freely described adjectives and merged synonyms to identify the 15 most frequently occurring adjective pairs for inclusion in the impression evaluation experiment. Since there were 8 pairs of evaluation items in the freely descriptions duplicate in the references, a total of 22 adjective pairs were ultimately used in the impression evaluation experiment, as shown in Table \ref{tab:1}.

\begin{table}
\centering
\caption{Impression Evaluation Items for Material Experience Based on Visual, Tactile and Smell Perception}
\label{tab:1}
\begin{tabularx}{\textwidth}{XXXX}
\toprule
\multicolumn{4}{l}{\textbf{Impression Evaluation Item}} \\ \midrule
Hard -- Soft          & Old -- New             & Dirty -- Clean        & Heavy -- Light         \\
Cold -- Warm          & Rustic -- Fancy        & Dark -- Bright        & Strong -- Weak         \\
Cloudy -- Clear       & Sticky -- Slippery     & Vulgar -- Elegant     & Sturdy -- Fragile      \\
Gloomy -- Shiny       & Cheap -- Luxury        & Artificial -- Natural & Orderly -- Disorderly  \\
Rough -- Smooth       & Traditional -- Modern  & Simple -- Complex     & Elasticity -- Plasticity \\
Monochrome -- Colourful & Bad Touch -- Good Touch &                    &                        \\ \bottomrule
\end{tabularx}
\end{table}

The purpose of this study was to investigate the art healing effects of material experiences by quantifying user feedback and exploring user-centered sustainability. To achieve this, five questions were designed, as shown in Table \ref{tab:2}, to explore the following aspects:

\begin{enumerate}
    \item Evaluate the material's ability to evoke either disgust or pleasure, thereby measuring the positive emotional connection between the material and the user. The Dislike - Like evaluation item can capture users' emotional responses in real time, revealing differences in product satisfaction and attractiveness across different user groups, helping to identify design issues, and is therefore widely used.
    \item Examine whether the material elicits feelings of anxiety or calmness, highlighting its emotion-regulating properties. Anxiety is a major psychological health issue that affects an individual's emotions, cognition, behavior, and physiological responses, while interventions targeting Anxiety - Relaxation evaluation item have shown significant effectiveness in improving mental health \cite{zhangEffectsArtTherapy2024}.
    \item Explore whether the material stimulates users' curiosity and creativity, emphasizing its cognitive and motivational effects. Creativity provides a "non-confrontational" means of expression for healing, allowing individuals to process inner conflicts in a safer and more flexible way. The Boring - Interesting is an important evaluation item for assessing creativity \cite{barnettHowArtsHeal2024, jean-berlucheCreativeExpressionMental2024}.
    \item Determine whether the material induces mental fatigue or revitalizes users, assessing its impact on mental energy levels. Fatigue is a complex subjective experience that encompasses multiple domains, including psychological, physical, and emotional aspects, and may arise from physical activity, mental activity, or emotional activity \cite{raizenSymptomBiologyFatigue2023}.
    \item Ascertain whether the material causes physiological or psychological discomfort. The World Health Organization (WHO) emphasizes the significant role of product comfort in enhancing user experience and treatment compliance \cite{maulaComfortEvaluationPhysiological2024}. At the same time, a comfortable state helps reduce physical stress, improve emotional stability, and enhance cognitive performance.
\end{enumerate}

\begin{table}[htbp]
\centering
\caption{Questionnaire for Exploring User-Centered Sustainability}
\label{tab:2}
\begin{tabularx}{\textwidth}{X}
\toprule
\textbf{\textit{Questions Related to Art Healing in Material Experience}} \\ \midrule
\textit{Does the material evoke a tendency toward repulsion or pleasure? (Dislike -- Like)} \\ \midrule
\textit{Does the material make you feel nervous or calm? (Anxiety -- Relaxation)} \\ \midrule
\textit{Does the material stimulate your desire to explore or creativity? (Boring -- Interesting)} \\ \midrule
\textit{Does the material make you feel mentally fatigued or energized? (Fatigue -- Vitality)} \\ \midrule
\textit{Does the material evoke physical or psychological discomfort? (Discomfort -- Comfort)} \\ \bottomrule
\end{tabularx}
\end{table}

\subsection{Data processing and analysis}
This study employed a 7-point Semantic Differential (SD) method to collect data through degree evaluations of paired descriptive adjectives. Specifically, a structured questionnaire was used to invite 10 participants to rate 28 stimuli, resulting in a total of 280 valid data. After data verification, it was confirmed that all samples had no missing values and demonstrated good integrity, making them directly suitable for subsequent statistical analysis.

This study employed IBM SPSS Statistics 27.0 to perform descriptive statistics, factor analysis, and hierarchical clustering, while the construction and validation of the structural equation model (SEM) were carried out using SPSS Amos 29.0. Additionally, Python (version 3.13.3) with the matplotlib library (version 3.10.3) was used for plotting the figures. A significance level of $p < 0.05$ was adopted for statistical significance, and all statistical tests were two-tailed to ensure the reliability and rigor of the results. The statistical significance levels were defined as follows: * denoted $0.01 \le p < 0.05$, ** denoted $0.001 \le p < 0.01$, and *** denoted $p < 0.001$.

\subsubsection{Exploratory Factor Analysis (EFA)}
In the factor analysis process, this study applied criteria for variable selection and model validation: Only common factors with eigenvalues greater than 1.0 were retained \cite{zwickComparisonFiveRules1986}, which effectively refined the core information of the original variables and ensured sufficient explanatory power. For the impression evaluation items, only those with a factor loading greater than 0.6 and a cross-loading less than 0.4 were included to ensure the uniqueness and independence of item attribution, and maintain the clarity and stability of the factor structure.

In the assessment of model fitting, multi-dimensional indicators were used to verify the validity of the factor analysis results: The Kaiser-Meyer-Olkin (KMO) measure of sampling adequacy exceeded 0.7, and Bartlett's Test of Sphericity yielded a statistically significant result (p < 0.05), collectively confirming the data's suitability for factor analysis. The cumulative variance explained by the extracted common factors reached greater than 60\%, demonstrating that the retained factors adequately captured the majority of variance in the original variables. 

\subsubsection{Structural Equation Model (SEM)}
Structural Equation Modeling (SEM), also known as covariance structure modeling or causal modeling, is a popular, powerful, and advanced statistical technique \cite{klinePrinciplesPracticeStructural2011} that integrates regression analysis and factor analysis to estimate structural equations involving latent variables \cite{bollenStructuralEquationsLatent1989}. The purpose of SEM is theory testing \cite{marcoulidesModernMethodsBusiness1998} and developing explanatory structural models of causal effects \cite{klinePrinciplesPracticeStructural2011}. This study employs the Maximum Likelihood Estimation (MLE) method within SEM to explore the transformation path from sustainable materials to user-centered sustainability based on material experience.

To assess the model's goodness-of-fit, seven well-established indicators commonly used in structural equation modeling were employed. The $\chi^2/df$ ratio measures the discrepancy between the observed and predicted covariance matrices. The Goodness-of-Fit Index (GFI) and Adjusted Goodness-of-Fit Index (AGFI) quantify the proportion of variance and covariance explained by the model, while the Comparative Fit Index (CFI) and Tucker-Lewis Index (TLI) compare the proposed model against a baseline. The Root Mean Square Error of Approximation (RMSEA) and Standardized Root Mean Square Residual (SRMR) measure the average differences between observed and predicted correlations \cite{iacobucciStructuralEquationsModeling2010, sahooStructuralEquationModeling2019}. The evaluation adhered to the following benchmarks: $\chi^2/df \le 3.00$; GFI, AGFI, and CFI $\ge 0.90$; TLI $\ge 0.95$; and RMSEA and SRMR $\le 0.08$, ensuring a rigorous yet practical assessment of model adequacy. For construct reliability and validity assessment, composite reliability (CR) values exceeded 0.7, meeting the threshold for acceptable internal consistency. Convergent validity was evaluated via the average variance extracted (AVE), with all factors achieving AVE values greater than 0.5, confirming their ability to effectively converge on the shared variance of the measured items \cite{fornellEvaluatingStructuralEquation1981}.

\subsubsection{Hierarchical Cluster}
The mean values of the impression evaluation items incorporated into the structural equation model were calculated, between-groups linkage as the clustering method and Squared Euclidean Distance as the measure of data similarity. This methodology enabled the quantitative assessment of similarities and differences in sensory properties of materials, presenting the correlations and distinctions between materials in a structured, visual manner.

\subsubsection{Factor Scores}
The mean scores of each sustainable material on the various factors were calculated and displayed in a three-dimensional space. Additionally, the pairwise relationships between factors were presented on a two-dimensional plane, and the correlations between the factors were computed. Meanwhile, t-tests were conducted comparing the scores of stimuli 1, which contained only the matrix material, with those of the other 27 sustainable materials across each factor. The number of stimuli showing significant increases, significant decreases, or no significant differences compared to stimulus 1 were then summarized.

\subsection{Participants and Experimental Environment}
There were 10 participants in the impression evaluation experiment, representing Tsinghua University, Beijing University of Chemical Technology, and Chiba University. The participants, aged between 23 and 35, included 5 males and 5 females. Each participant had at least 4 years of education in design or materials, holding degrees in engineering, science, or art. Among them, 5 participants hold bachelor's degrees, 3 hold master's degrees, and 2 hold doctoral degrees. These participants not only have extensive experience in product and material development but also possess research backgrounds in industrial design, user experience, material design, Kansei engineering, and CMF-related fields. Compared to regular users, they have a deeper understanding of materials and more refined perceptual abilities. Given their professional expertise, we believe they offer more valuable insights into interpreting material experience and its healing effects. Additionally, their diverse knowledge across multiple fields enables a more comprehensive understanding of the issue. Therefore, we chose to conduct the experiment with a small group of specialized participants, aiming to explore the trends between materials and perception.

Prior to the start of the experiment, each evaluation item was explained sufficiently to ensure that participants were consistent in their understanding of the adjectives. Participants were allowed to experience all of the stimuli freely to ensure that they fully understood the differences between the individual stimuli. There was no time limit on the experience of the stimuli, and participants were allowed to experience the materials based on visual, tactile and smell sensations during the evaluation experiments. The experimental conditions for the impression evaluation were 25\textdegree C, 30\% humidity, 300--500\,lx of illumination, and a color temperature of 4000--4500\,K.

\section{Results and Discussions}
\subsection{Exploratory Factor Analysis}
The Kaiser-Meyer-Olkin (KMO) measure of sampling adequacy is 0.925, and Bartlett's Test of Sphericity yields a p-value less than 0.001, indicating statistical significance. The factor analysis revealed three factors with eigenvalues greater than 1, which collectively explain 70.052\% of the total variance. Among the initial 22 evaluation items, two items, "Cold - Warm" and "Traditional - Modern", were removed due to factor loadings below 0.6. The results are shown in Table \ref{tab:3}.

\begin{table}[htbp]
\centering
\caption{Pattern Matrix from Factor Analysis Using Maximum Likelihood Estimation with Promax Rotation}
\label{tab:3}
\begin{tabularx}{\textwidth}{
    >{\hsize=2.4\hsize}L 
    >{\hsize=0.65\hsize}C 
    >{\hsize=0.65\hsize}C 
    >{\hsize=0.65\hsize}C 
    >{\hsize=0.65\hsize}C
} 
\toprule
\multicolumn{5}{l}{\textbf{\textit{Factor Loading and Variance Explained}}} \\ \midrule
\multirow{2}{*}{Evaluation Item} & \multicolumn{3}{c}{Component} & \multirow{2}{*}{Extraction} \\ 
\cmidrule(lr){2-4}
& Aesthetic & Intrinsic & Physical &  \\ \midrule

Artificial -- Natural      & 0.906  & -0.104 & -0.017 & 0.771 \\
Vulgar -- Elegant          & 0.905  & 0.119  & -0.005 & 0.885 \\
Simple -- Complex          & 0.862  & -0.160 & 0.000  & 0.696 \\
Monochrome -- Colourful    & 0.813  & -0.012 & -0.098 & 0.602 \\
Rustic -- Fancy            & 0.811  & 0.051  & 0.042  & 0.714 \\
Cheap -- Luxury            & 0.788  & 0.075  & -0.046 & 0.627 \\
Orderly -- Disorderly      & 0.723  & -0.145 & 0.059  & 0.519 \\
Bad Touch -- Good Touch    & 0.654  & 0.239  & 0.104  & 0.656 \\
\midrule
Cloudy -- Clear            & -0.055 & 0.880  & -0.004 & 0.750 \\
Gloomy -- Shiny            & -0.074 & 0.844  & -0.052 & 0.652 \\
Dark -- Bright             & -0.005 & 0.827  & -0.018 & 0.669 \\
Dirty -- Clean             & -0.007 & 0.827  & 0.018  & 0.694 \\
Old -- New                 & 0.022  & 0.819  & -0.060 & 0.640 \\
Intense -- Mild            & 0.040  & 0.707  & 0.083  & 0.577 \\
\midrule
Sturdy -- Fragile          & -0.065 & -0.158 & 0.919  & 0.703 \\
Heavy -- Light             & 0.075  & -0.044 & 0.795  & 0.655 \\
Rough -- Smooth            & -0.031 & 0.174  & 0.734  & 0.660 \\
Strong -- Weak             & 0.023  & -0.099 & 0.699  & 0.450 \\
Plasticity -- Elasticity   & -0.048 & 0.083  & 0.687  & 0.502 \\
Sticky -- Slippery         & 0.032  & 0.123  & 0.668  & 0.554 \\
\midrule
Eigenvalue                 & 7.810  & 3.865  & 2.335  & --- \\
Cumulative (\%)            & 39.050 & 19.325 & 11.677 & --- \\
Cumulative Contribution Ratio (\%)  & 39.050 & 58.375 & 70.052 & --- \\ \bottomrule
\end{tabularx}
\end{table}

The results of the factor analysis show that the first factor includes 8 items, which can explain 39.050\% of the total variance, and the Cronbach's alpha coefficient reaches 0.937, which is extremely strong in internal consistency. This factor focuses on the aesthetic, morphological and textural dimensions, named as "Aesthetics" property. The second factor contains 6 items, explaining 19.325\% of the total variance, with a Cronbach's alpha of 0.921, and its content mainly involves the material's light reflectivity, surface cleanliness, and the change of the state under the action of time, which are the intrinsic attributes of the material formed in the natural environment, independent of any temporary influence from external conditions, and therefore it is named as "Intrinsic" property. The third factor also comprises 6 items, explaining 11.677\% of the total variance, with a Cronbach's alpha of 0.885. It is directly associated with the material's physical and mechanical property—such as hardness, weight, and surface roughness—which can be measured through physical experiments and direct observation. This factor primarily reflects the material's physical characteristics and is therefore named "Physical" property.

\subsection{Structural Equation Model}
\begin{figure}
\centering
\includegraphics[width=1\textwidth]{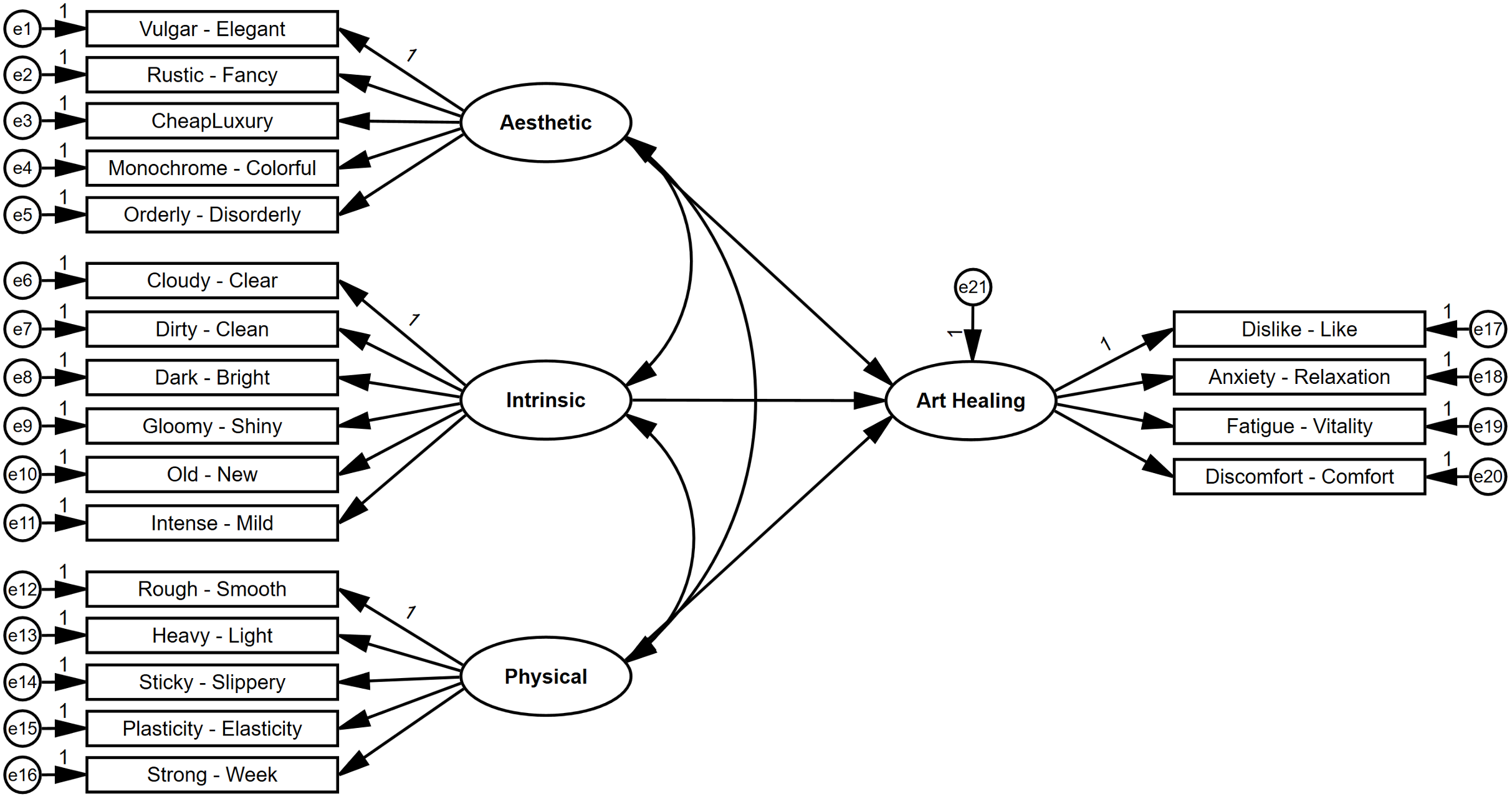}
\caption{Exploring the Transition from Sustainable Materials to User-Centered Sustainability through Art Healing: Structural Equation Modeling Path Diagram}
\label{fig:3}
\end{figure}

Based on the results of exploratory factor analysis, we constructed a structural equation model and optimized the model fit through adjustments. The final model is presented in Figure \ref{fig:3}. Table \ref{tab:4} summarizes the model fit indices. The analysis shows that indices such as $\chi^2/df$, GFI, CFI, TLI, RMSEA, and SRMR all demonstrate an excellent model fit. Although the AGFI does not fully reach 0.9, it is very close and falls within an acceptable range. Collectively, these indicators confirm that the model achieves a good fit and effectively represents the internal relationships among variables. Furthermore, this result also demonstrates that the theoretical framework constructed in this study possesses high accuracy and scientific validity.

\begin{table}[htbp]
\centering
\caption{Model Fit Indices and Model Evaluation}
\label{tab:4}
\begin{tabularx}{\textwidth}{lCCCCCCC}
\toprule
\multicolumn{8}{l}{\textbf{\textit{Model Fit Indices}}} \\ \midrule
& $\chi^2/df$ & GFI & AGFI & CFI & TLI & RMSEA & SRMR \\ \midrule
Cutoff value     & $< 3.00$ & $> 0.900$ & $> 0.900$ & $> 0.900$ & $> 0.950$ & $< 0.080$ & $< 0.080$ \\
Analysis Results & 1.647    & 0.910    & 0.885    & 0.972    & 0.968    & 0.048    & 0.050    \\
Model Evaluation & Good fit & Good fit & Accept   & Good fit & Good fit & Good fit & Good fit \\ \bottomrule
\end{tabularx}
\end{table}

Table \ref{tab:5} and \ref{tab:6} report the parameter estimates of the structural equation model, as well as the composite reliability (CR), average variance extracted (AVE), and covariance matrix between factors, respectively. The results indicate that the CR values for all constructs exceed 0.7, and the AVE values surpass 0.5, suggesting that the model possesses satisfactory internal consistency reliability and convergent validity. In addition, the square roots of the AVE values for each construct are greater than their correlation coefficients with other constructs (Table \ref{tab:7}), indicating that the model also demonstrates acceptable discriminant validity.

\begin{table}
\centering
\caption{Structural Equation Modeling Parameter Estimates}
\label{tab:5}
\begin{tabularx}{\textwidth}{l l C C C C C C}
\toprule
\multicolumn{8}{l}{\textbf{\textit{Structural Equation Modeling Parameters}}} \\ \midrule
Factors & Item & Un-Std. & S.E. & t-value & P & Std. & SMC \\ \midrule
\multirow{5}{*}{Aesthetic} 
& Vulgar -- Elegant      & 1.000 &       &        &       & 0.944 & 0.891 \\
& Rustic -- Fancy        & 0.936 & 0.041 & 22.761 & ***   & 0.865 & 0.748 \\
& Cheap -- Luxury        & 0.660 & 0.036 & 18.514 & ***   & 0.790 & 0.624 \\
& Monochrome -- Colourful & 0.756 & 0.045 & 16.615 & ***   & 0.747 & 0.558 \\
& Orderly -- Disorderly  & 0.687 & 0.050 & 13.717 & ***   & 0.669 & 0.448 \\ \midrule
\multirow{6}{*}{Intrinsic} 
& Cloudy -- Clear        & 1.000 &       &        &       & 0.867 & 0.752 \\
& Dirty -- Clean         & 0.974 & 0.055 & 17.859 & ***   & 0.833 & 0.694 \\
& Dark -- Bright         & 0.969 & 0.056 & 17.298 & ***   & 0.817 & 0.667 \\
& Gloomy -- Shiny        & 0.877 & 0.052 & 16.788 & ***   & 0.803 & 0.645 \\
& Old -- New             & 0.903 & 0.055 & 16.384 & ***   & 0.791 & 0.626 \\
& Intense -- Mild        & 0.848 & 0.055 & 15.417 & ***   & 0.762 & 0.581 \\ \midrule
\multirow{5}{*}{Physical} 
& Rough -- Smooth        & 1.000 &       &        &       & 0.855 & 0.731 \\
& Sticky -- Slippery     & 1.068 & 0.072 & 14.764 & ***   & 0.784 & 0.615 \\
& Heavy -- Light         & 0.996 & 0.071 & 14.080 & ***   & 0.757 & 0.573 \\
& Plasticity -- Elasticity & 0.960 & 0.073 & 13.203 & ***   & 0.721 & 0.520 \\
& Strong -- Weak         & 0.751 & 0.076 & 9.849  & ***   & 0.571 & 0.326 \\ \midrule
\multirow{4}{*}{Art Healing} 
& Dislike -- Like        & 1.000 &       &        &       & 0.889 & 0.790 \\
& Anxiety -- Relaxation  & 1.103 & 0.056 & 19.787 & ***   & 0.858 & 0.736 \\
& Fatigue -- Vitality    & 1.001 & 0.052 & 19.349 & ***   & 0.848 & 0.719 \\
& Discomfort -- Comfort  & 0.930 & 0.052 & 17.827 & ***   & 0.812 & 0.659 \\ \bottomrule
\end{tabularx}
\end{table}

\begin{table}
\centering
\caption{Composite Reliability (CR), Average Variance Extracted (AVE), and Correlation Coefficients}
\label{tab:6}
\begin{tabularx}{\textwidth}{l CCCCCC}
\toprule
\multicolumn{7}{l}{\textbf{\textit{CR, AVE and Correlation Coefficients Matrix}}} \\ \midrule
Factor & CR & AVE & Aesthetic & Intrinsic & Physical & Art Healing \\ \midrule
Aesthetic   & 0.903 & 0.654 & 0.809 &       &       &       \\
Intrinsic   & 0.921 & 0.661 & 0.310 & 0.813 &       &       \\
Physical    & 0.859 & 0.553 & 0.443 & 0.497 & 0.744 &       \\
Art Healing & 0.914 & 0.726 & 0.785 & 0.557 & 0.408 & 0.852 \\ \bottomrule
\end{tabularx}
\end{table}

\begin{table}
\centering
\caption{The Impact of Experiential Property of Sustainable Materials on User-Centered Sustainability}
\label{tab:7}
\begin{tabularx}{\textwidth}{l c l CCCCC}
\toprule
\multicolumn{8}{l}{\textbf{\textit{Path coefficients}}} \\ \midrule
\multicolumn{3}{l}{Path Analyses} & Un-Std. & S.E. & t-value & P & Std. \\ \midrule
Aesthetic & $\rightarrow$ & \multirow{3}{*}{Art Healing} & 0.547  & 0.039 & 14.019 & ***   & 0.706  \\
Intrinsic & $\rightarrow$ &                             & 0.413  & 0.050 & 8.177  & ***   & 0.408  \\
Physical  & $\rightarrow$ &                             & -0.128 & 0.062 & -2.068 & 0.039 & -0.107 \\ \bottomrule
\end{tabularx}
\end{table}

Table \ref{tab:7} reports the path coefficients among factors. The results indicate that the standardized regression coefficients of "Aesthetic", "Intrinsic", and "Physical" property on "Art Healing" are 0.706***, 0.408***, and 0.107*, respectively. A further analysis reveals that the "Aesthetic" property exerts the strongest influence on "Art Healing", followed by the "Intrinsic" property, while the contribution of the "Physical" property is comparatively limited.

\subsection{Hierarchical Cluster}
\begin{figure}
\centering
\includegraphics[width=1\textwidth]{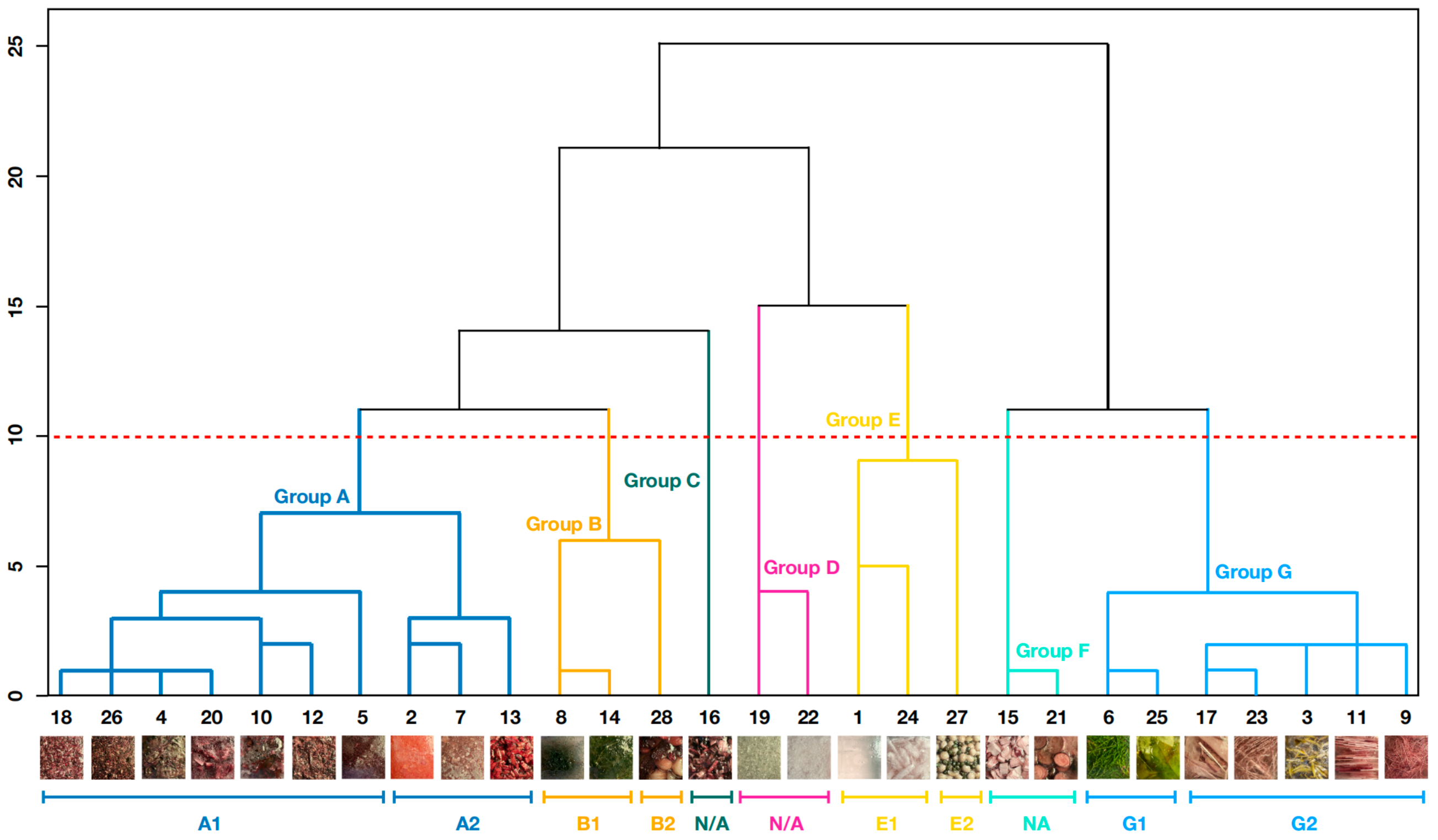}
\caption{Results of hierarchical cluster analysis based on intergroup linkage method with Squared Euclidean Distance, based on the characteristics of the three factors, a total of seven clusters and eleven sub-clusters were identified. Cluster A consists of wood and fruit residues (fine particles, coarse particles). Cluster B includes seaweed and dried fruits and nuts. Cluster C comprises traditional Chinese medicine residues. Cluster D contains leather. Cluster E includes unfilled materials, bagasse, soybeans, and mung beans. Cluster F consists of wood (with spatial structure). Cluster G includes fibrous and flaky plants}
\label{fig:4}
\end{figure}

The results of the hierarchical cluster analysis are shown in Figure \ref{fig:4}. Based on the dendrogram cut-off line, the 28 stimuli are categorized into seven major clusters (Groups A–G), with each group further divisible into subclusters based on finer-grained characteristics, as detailed below:

\begin{itemize}
    \item Group A: Group A consists of materials derived from renewable sources, such as plant-based residues. Data analysis shows that the stimuli in this group scored slightly below the neutral point (scale midpoint = 4) across the three factors of "Aesthetic", "Intrinsic", and "Physical" properties. In combination with interview findings, these environmentally friendly materials are perceived as relatively familiar in daily life, with sensory characteristics of the composites closely resembling those of the filler materials. A further subdivision of Group A reveals that, compared to Sub-Group A1, materials in Sub-Group A2 display more distinctive features. For example, Stimuli 2 and 7, made from different types of fruit residues, emit a natural aroma and present a unique sensory experience; Stimulus 13 stands out for its vibrant color. These stimuli received higher overall ratings across all factors compared to those in Sub-Group A1. Although participants could clearly perceive the sustainability of the materials, their sensory qualities were relatively weak.
    \item Group B: Group B consists of two subgroups: Subgroup B1 uses seaweed as the filler, while Subgroup B2 is composite made from a mixture of dried fruits and nuts. These two materials scored lower on the other four items within the "Aesthetic" property but received higher scores on the "Orderly - Disorderly" item. In addition, they received higher scores in the "Physical" properties, with participants generally perceiving their surfaces as smooth in texture. However, due to their dull coloration and the chaotic structure of the fillers, they failed to evoke a desire for continued use.
    \item Group C: In Group C, only one stimulus is included—Stimulus 16. The filler used in this stimulus is highly distinctive, being made from residues of traditional Chinese medicine. It received extremely low scores across all factors. This finding is further supported by participant interviews, in which all participants explicitly reported an extremely poor sensory experience, including visual, tactile, and smell aspects. We hypothesize that this negative cognition may be closely associated with painful memories of taking medicine.
    \item Group D: Group D consists of Stimulus 19 and 22, both of which use leather filler. Compared to other sustainable materials, these are reprocessed products with lower levels of sustainability and environmental friendliness. The materials in this group are characterized by lower scores in the "Aesthetic" property and higher scores in the "Intrinsic" property. This may be because natural materials have more uncontrollable random features, while leather products processed twice have more uniform appearances but lose the randomness of natural materials.
    \item Group E: Group E consists of Stimuli 1, 24, and 27. Statistical results show that these stimuli scored highly in the "Intrinsic" and "Physical" properties, specifically exhibiting strong cleanliness, bright luster, soft touch, weightlessness, and high elasticity. Participants reported that this group of materials evoked a sense of simplicity and freshness, which in turn enhanced their willingness for continued use. The difference between Subgroup E1 and Subgroup E2 lies in the latter's addition of a large number of soybeans and mung beans, causing the material to partially lose elasticity and increase in weight.
    \item Group F: Group F includes only Stimulus 15 and 21. The stimuli in this group use wood as the filler. However, unlike the powdered fillers in Group A, the fillers in Group F are wood blocks. Statistical results indicate that the materials in this group scored highly in the "Aesthetic" property. Participants generally perceived the stimuli as having strong healing, particularly due to the sense of spatiality created by the wood blocks within the matrix, which significantly enhanced the aesthetic appeal of the materials. Although they scored lower in the "Physical" property—being perceived as heavy and rough—this had only a limited impact on their healing effects.
    \item Group G: Group G consists of two subgroups, comprising a total of seven stimuli. The stimuli in this group are primarily derived from intact plant materials and exhibit fibrous or flaky morphological characteristics. Participants widely acknowledged that the experiential qualities of these materials contributed to a sense of emotional healing. Data also indicated that the stimuli in this group received high scores across all three factors. Specifically, Subgroup G1 outperformed Subgroup G2 in all factors, mainly due to its vibrant green appearance. In contrast, the less vivid coloration of Subgroup G2 was identified as the primary reason for its slightly lower ratings. Participant interviews further revealed that the stimuli in Subgroup G1 were more highly regarded, with particular emphasis on the strong healing sensation evoked by the green color and a preference for the material's tactile quality. In addition, we observed that intact, unprocessed filler allowed participants to perceive the original vitality of the plants. Whether in fibrous or flaky shape, these fillers in the original form of sustainable materials make a positive contribution to the material experience.
\end{itemize}

Through hierarchical cluster analysis, the multidimensional characteristics of the stimuli can be more clearly deconstructed, thereby providing empirical support for validating the theoretical framework of the structural equation model. This analytical process serves to inspire designers to further explore the latent potential of sustainable materials. With design approaches, sustainable materials are transformed into art carriers with healing functions, realising the precise release of art's healing effects. Building upon this foundation, the study further advocates for a user-centered approach to sustainable design, aiming to construct a dynamically balanced symbiotic relationship among humans, materials, and the environment.

\subsection{Factor Score}
\begin{figure}
\centering
\includegraphics[width=0.9\textwidth]{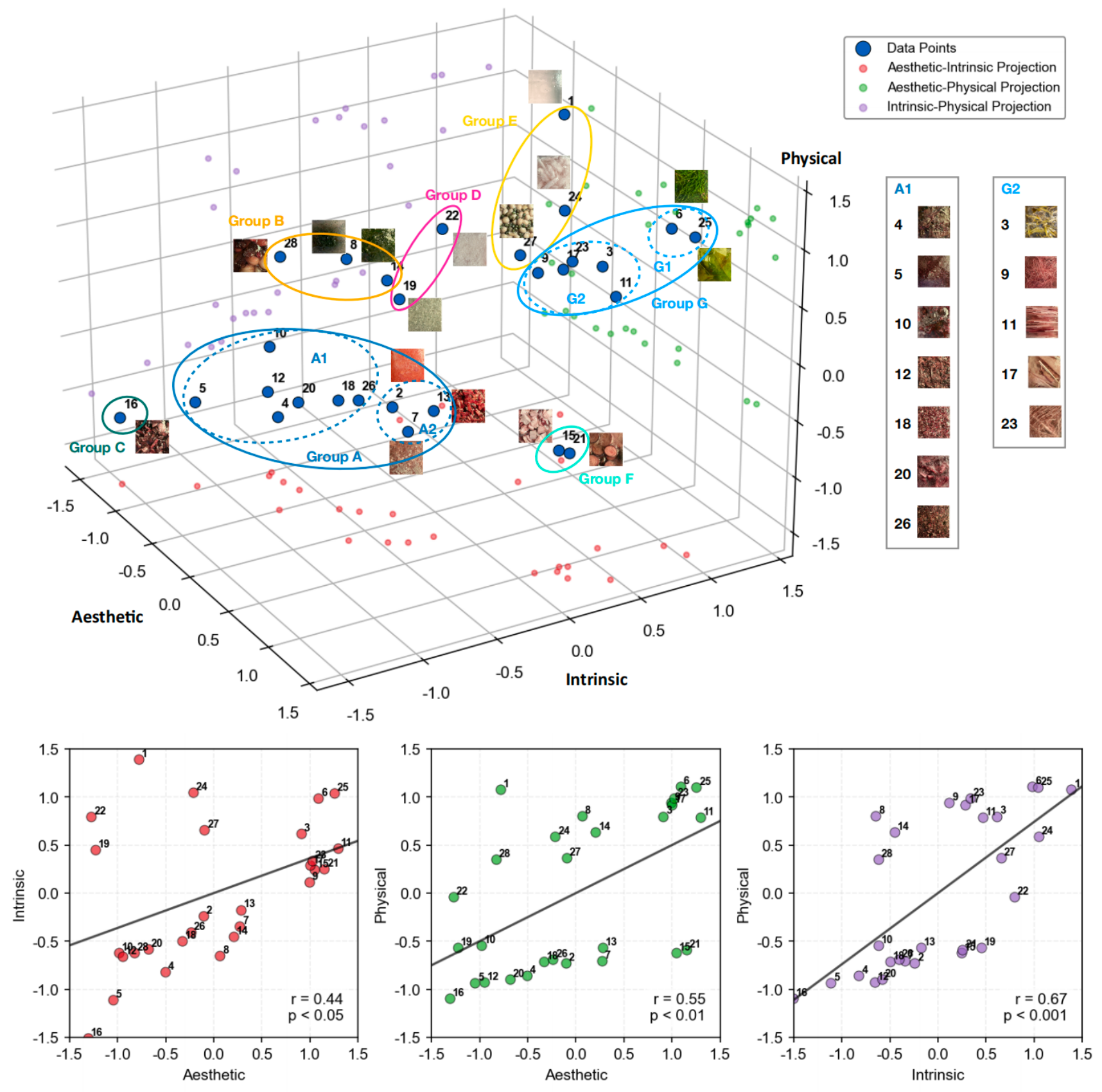}
\caption{The upper section shows the mean factor score distribution of the 28 stimuli across the three factors, as well as their projections onto two-dimensional planes formed by each pair of factors. The resulting from the hierarchical cluster analysis is also labeled. The lower section presents the results of the correlation analysis between each pair of factors and the distribution of the stimuli}
\label{fig:5}
\end{figure}

Based on the evaluation items in the structural equation model, factor scores for each stimulus across the three factors were calculated, and their mean values were obtained through statistical analysis. Figure \ref{fig:5} shows the distribution of scores on the three factors for each stimulus. The scores were also projected onto planes formed by each pair of factors, and the correlation coefficients between factors were calculated. In addition, the stimuli were labeled in the figure according to the results of the hierarchical cluster analysis.

The results of the correlation analysis indicate that the correlation coefficient between the "Aesthetic" and "Intrinsic" properties is 0.44* (p < 0.05), showing statistical significance. The correlation between the "Aesthetic" and "Physical" properties is 0.55** (p < 0.01), but it does not reach a significant level. The correlation between the "Intrinsic" and "Physical" properties is 0.67*** (p < 0.001), indicating a high level of significance. These findings suggest that there may be interactions among the factors. We infer that, during material experience, interactions may occur between multisensory and multidimensional perceptions, and such interactions may influence the overall material experience as well as the resulting healing effects. Therefore, researchers and designers can expand and enhance the perceptual properties of materials by skilfully utilising their interactions.

\begin{figure}
\centering
\includegraphics[width=1\textwidth]{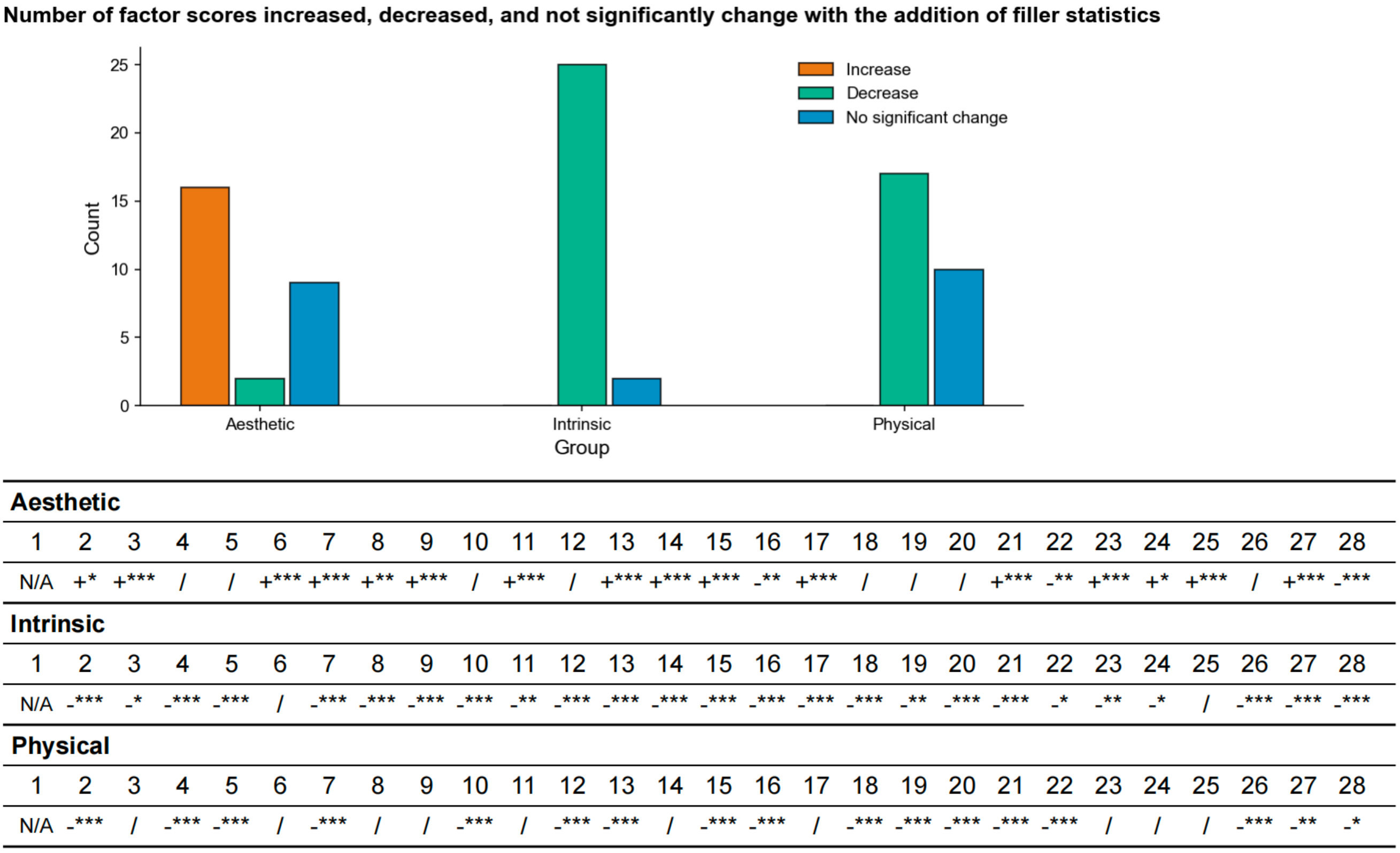}
\caption{Statistics on the Number of Stimulus Changes in Each Factor After Adding Fillers}
\label{fig:6}
\end{figure}

To further investigate the impact of fillers on material experience, a t-test was conducted between the stimuli containing only the matrix and the sustainable composite materials. Figure \ref{fig:6} summarizes the changes in factor scores across each factor after the addition of sustainable fillers.

In the quantitative assessment of the "Aesthetic" property, the standardized z-score of Stimulus 1—used as the baseline reference—is -0.728, indicating a relatively low aesthetic level. This score suggests that the stimulus is perceived to exhibit characteristics such as vulgarity, rusticity, cheapness, monochromaticity, and orderliness. Following the incorporation of filler, 16 stimuli demonstrated significant improvements in aesthetic scores, while only 2 stimuli showed a decrease, and 9 stimuli exhibited no statistically significant change. Results from the SEM further indicate that the "Aesthetic" property holds the highest influence weight on the effectiveness of art healing among all factors. These findings highlight that selecting filler of significantly enhancing aesthetic value is not only a key strategy for optimizing the healing efficacy of materials, but also a critical step toward advancing the development of sustainable materials—from a traditional function-oriented paradigm to a user- centered approach—thereby promoting integrated social and ecological sustainability.

In the "Intrinsic" property, baseline stimulus 1 had a high standardized score of 1.389, indicating that its perceptual characteristics were clarity, brightness, and mildness. After the addition of filler, none of the sustainable material exhibited score improvements in this factor—25 stimuli showed significant declines, while only 2 stimuli demonstrated no statistically significant change. Although "Intrinsic" property function as secondary factors influencing the outcomes of art healing, variations in this factor still exert a non-negligible impact. Therefore, we consider that effective management and minimisation of negative changes in this factor is essential to maintain healing efficacy.

In the "Physical" property, the standardized score of the baseline Stimulus 1 is 1.081, indicating that the stimulus exhibits characteristics of smoothness and elasticity. After the addition of filler, 17 sustainable materials exhibited increased roughness and enhanced plasticity, while the remaining 10 sustainable materials showed no significant changes. Notably, fluctuations in the physical properties did not have a significant impact on the effectiveness of art healing. This result highlights a key distinction from traditional material engineering: in the field of design, mechanical properties are not the primary concern. Instead, the key lies in fully leveraging the perceptual property of sustainable material to enhance the effects of art healing, thereby achieving user-centered sustainable. Therefore, when developing user-centered sustainable materials or products, greater emphasis should be placed on the design of aesthetic experience and psychological perception.

\section{Conclusions}
This study establishes a sustainable materials system using hydrogels and various sustainable fillers to explore how perceived material properties critically affect the outcomes of art healing. And highlights the essential role of sensory experience in the user-centered sustainable. The findings show that "Aesthetic" property possess not only strong explanatory power within the SEM but also represent the most emotionally evocative in participants' interviews. In contrast, "Intrinsic" property exerts a moderate effect, while "Physical" property shows relatively limited effect.

Expanding on these insights, the study demonstrates that the morphology and color characteristics of sustainable material significantly influence users' experience of art healing. Sustainable material featuring natural form and vibrant colour are more effective in evoking positive emotions and fostering a sense of psychological connection with users. This suggests that future sustainable material design should move beyond mere environmental and functional concerns to also prioritize the emotional responses materials can evoke in human-material interactions.

Furthermore, the study emphasizes the crucial role of multisensory integration in shaping material experiences. The interaction between vision, touch and smell stimuli can enhance or dampen users' overall perception of a material, indicating that cross-modal sensory interactions serve as a moderator in material experience. This provides valuable theoretical insights and practical strategies for designers aiming to optimize material experience through a multimodal approach.

Additionally, this study highlights a fundamental distinction between material-driven design and materials engineering. While the latter primarily concentrates on optimizing physical performance, the former places greater emphasis on the user's perceptual and emotional experience. By utilizing fillers not only to enhance functionality but also to enrich emotional appeal, design-oriented approaches aim to imbue materials with healing qualities and emotional resonance. This strategy can elevate psychological value, deepen user attachment, and encourage sustained engagement, thereby extending product lifespans and fostering a harmonious synergy between sustainable materials and user-centered sustainability.

In summary, user-centered sustainability should transcend their physical properties of eco-friendliness and recyclability to address the deeper question of how individuals emotionally connect with materials. By enhancing aesthetic expression, positive psychological experiences and adopting multi-sensory design strategies, future material innovation can shift from an environmentally driven paradigm to an emotionally and humanistically valued paradigm, opening up new avenues for health-orientated and emotionally engaging product and system design.

\subsection{Theoretical Implications}
This study proposes a research framework for material experience, providing designers with a systematic and structured methodological model to understand material properties, and introduces it into the interdisciplinary context of art healing. It not only expands the boundaries of interdisciplinary applications among design, materials science, and psychology, but also offers interdisciplinary theoretical support for user-centered material innovation in the context of sustainable design, demonstrating the integrative value of design disciplines in addressing complex social challenges.

\subsection{Limitations}
In this study, hydrogel was chosen as a matrix because of its potential for a wide range of applications in the field of sustainable design, as well as its environmental friendliness and bio-solubility. However, the insufficient thermal stability of hydrogels at room temperature has an effect on the material experience. To address this issue, the study used an experimental protocol where the evaluation was completed within 30 minutes of removal from refrigeration.

Notably, given the small sample size design of this study, the quality of participants' evaluations became a critical consideration. To this end, all participants in this experiment had professional backgrounds. To enhance the generalizability of the research model, follow-up studies will expand the sample size and incorporate diverse data from non-professional groups.

\section{Statements and Declarations}
    The authors have no competing interests to declare that are relevant to the content of this article.

\bibliographystyle{unsrt}
\bibliography{references}






\end{document}